\documentclass[preprint,12pt,a4paper]{elsarticle} 
\usepackage{natbib}
\usepackage{amsmath}
\usepackage{amssymb}
\usepackage{graphics}
\usepackage{setspace}
\usepackage{hyperref}
\usepackage{epsfig}
\ifx\pdfoutput\undefined
 \usepackage{graphicx}
 \else
 \usepackage{graphicx}
 \usepackage{epstopdf}
\fi

\begin{document}
\title{Stochastic Boundary Conditions for Molecular Dynamics Simulations}
\author[rvt]{Manamohan Prusty}
\ead{mprusty@ntu.edu.sg}
\author[rvt]{Jia Ning Leaw}
\ead{jnleaw@ntu.edu.sg}
\author[xx]{Shang Shan Chong}
\author[rvt]{Siew Ann Cheong}
\ead{cheongsa@ntu.edu.sg}
\address[rvt]{Division of Physics and Applied Physics,
School of Physical and Mathematical Sciences,
Nanyang Technological University,
21 Nanyang Link, Singapore 637371, Republic of Singapore}
\address[xx]{8 Ellington Square, Singapore 568919, Republic of Singapore}
\begin{abstract}

In this paper we develop a stochastic boundary conditions (SBC) for event-driven molecular dynamics simulations of a finite volume embedded within an infinite environment. In this method, we first collect the statistics of injection/ejection events in periodic boundary conditions (PBC). Once sufficient statistics are collected, we remove the PBC and turn on the SBC. In the SBC simulations, we allow particles leaving the system to be truly ejected from the simulation, and randomly inject particles at the boundaries by resampling from the injection/ejection statistics collected from the current or previous simulations. With the SBC, we can measure thermodynamic quantities within the grand canonical ensemble, based on the particle number and energy fluctuations. To demonstrate how useful the SBC algorithm is, we simulated a hard disk gas and measured the pair distribution function, the compressibility and the specific heat, comparing them against literature values.
\end{abstract}
\maketitle
\section{Introduction}
Molecular dynamics is an important tool for understanding material as well
as thermodynamics behaviour of a system. It solves Newton's laws of
motion for the trajectories of the particles inside a given
system~\cite{AT-91,FS-02}. At present, state-of-the art MD simulations can be
carried out for $10^{10} - 10^{11}$ particles over $10^3$ nodes \cite{GK-05,R-06}. 
However, this is still many orders of magnitude away from truly macroscopic 
systems, which have $\mathcal{O}(10^{23})$ microscopic variables. Depending on 
which physical quantities we want to measure from the simulations, finite size 
effects can be important. These can arise as a result of the finite number of particles simulated, and also from the presence of boundaries. The most common way to deal with MD simulations of finite systems is to impose periodic boundary conditions (PBC)~\cite{AT-91,FS-02,R-04,H-92}. 
With PBC, the primary simulation box is surrounded by images of itself, and whenever 
a particle exits the simulation box, an identical particle reenters the box through 
the opposite end. Thus the ejection of particles is correlated  with the injection 
of another particle at the opposite end. This is very different from what we expect of an observation windows embedded in an infinite system. 

Earlier attempts to remove the artificial PBC correlations can be classified into four categories. The first category is based on the concept of a heat bath. In Cicotti and Tenenbaum's simulation, particles are injected as in PBC, but with different velocities sampled from a Maxwell-Boltzmann distribution~\cite{CT-80, CTG-82}. This method breaks the correlation of momenta between ejected and injected particles, but position correlation still remains. In contrast, the stochastic changes in momenta occur throughout the simulation volume in Andersen's method~\cite{A-80}. Random particles are chosen to suffer stochastic collisions with time intervals between successive collisions sampled from a Poisson distribution, and the resultant velocities sampled from a Maxwell-Boltzmann distribution. Another method based on the concept of heat bath was proposed by Berendsen~\cite{BPvDH-84}. In this method, the simulated system is coupled to a heat bath at constant temperature. Heat flow in and out of the system at a rate proportional to the difference between its kinetic temperature and the heat bath temperature. The velocities of particles in the system are then rescaled according to the heat that flows in. Except for the method by Cicotti and Tenenbaum, these heat bath methods are designed for simulating only equilibrium quantities, and not dynamic quantities. 

The second category consists of grand canonical molecular dynamics (GCMD) simulations. Cagin and Pettit~\cite{CP-91mp,CP-91ms}
developed a deterministic GCMD by writing down the Lagrangian of the system with the number of particles $n$ as one of the continuous variables. The fractional part of $n$ is then introduced as a fractional particle, while the integer part of $n$ represents full particles. The rate of change of the number of particles $\dot{n}$ is then derived from the Lagrangian, along with the rates of change of momenta, as the equations of motion of the system. When the fractional particle becomes a
full one, a new particle is added to the system; and when the fractional particle decreases to zero, it is 
deleted and another particle is chosen
as the fractional particle. To make the simulated dynamics as smooth as possible, the addition and the deletion of the particles are done at the region where the potential energy is the closest to the previously added or deleted particle. Newly added particles are assigned zero velocities. Unlike in a real system where exchange of particles takes
place only at the boundaries, the removal and addition of new particles in Cagin and Pettit's approach takes place within the interior of the system. Although it provides a straightforward simulation of grand canonical ensembles, 
it is difficult to use the method to study systems where particle exchange at the boundary is 
important e.g., gas exchange between two media. Another method, described by Heffelfinger 
and Van Swol~\cite{HS-94}, provides an answer to this problem. In their control
volume grand canonical molecular dynamics (CV-GCMD) method, the simulation system is surrounded by control volumes subject to PBC. Measurements are done only within the simulation 
box, even though a larger volume is simulated.

The third category introduces a stochastic boundary region surrounding the simulation volume. In the method developed by Berkowitz and McCammon \cite{MBJM-82}, the simulation system is divided into three regions, namely the simulation region, the bath region and the reservoir region. The simulation region consists of a central particle, and particles within a certain range from this central particle. Outside this simulation region, the bath region forms a shell that encloses the simulation region. The outermost region is the reservoir region. The particles in the simulation and bath regions follow MD and Langevin dynamics respectively, while particles in the reservoir are held fixed. When calculating forces acting on particles in the first two regions, the potentials arising from particles in all three regions are considered. Brooks and Karplus \cite{CLBMK-83} modified the method by replacing the reservoir region with a boundary region. The boundary region contains no particles, but generates a potential based on an average structure inferred from the radial distribution function inside the simulation region. To prevent particles in the bath region from venturing into the boundary region, and in effect changing the average structure within the boundary region, the whole boundary region is made repulsive. Boorks, Karplus and Brunger later applied the method to simulate water \cite{ABCLBMK-84} and proteins \cite{BBK-85}.

The forth category involves multiscale simulations, whereby the system is simulated using a combination of MD and a continuum method~\cite{XB-04, SCM-02, BOK-95}. The simulation system is separated mainly into three domains, viz. a molecular domain, a continuum domain and a bridging domain. The Lagrangians of the molecular and continuum domains are written down and their equations of motions are derived. Within the bridging domain, the Lagrangian is written as a linear combination of those of the other two domains. At each point in this bridging domain, molecular and continuum displacements are constrained to be the same. The constrained equations of motion are then obtained using the Lagrange multiplier method. While the multiscale approach is very appealing, there is yet no systematic study on how sensitive the simulation results are to different choices of the three domain sizes. There are also no serious efforts to determine whether the continuum Lagrangian is truly compatible with the atomistic Lagrangian, i.e. can we derive the continuum Lagrangian through coarse graining the atomistic Lagrangian? In view of these open questions, the multiscalse simulation approach can at best be a complement, but not a substitute for fully-atomistic, first-principle simulations.

To more accurately simulate a finite observation window embedded in an infinite system, we propose using a hierarchy of stochastic boundary conditions (SBC) that is closest in spirit to the stochastic boundaries developed by Berkowitz and McCammon \cite{MBJM-82}, and extended by Brooks and Karplus \cite{CLBMK-83}. In this method, which is based on the concept of resampling, stochastic events at the boundaries of the system are sampled from the statistics collected within the system itself, so that there is never the need to simulate a larger supersystem. The ejected particles are simply deleted, and not re-injected like in PBC. While particles are ejected, new particles are injected with random velocities at random positions along the boundaries, and at a rate that is consistent with the simulation volume being embedded in an infinite system at equilibrium temperature $T$. The approach is hierarchical. To get greater accuracy, we simply go to higher order. The detailed implementation of the SBC would depend on whether we are simulating a gas, a solid or a liquid, and also on whether we are working with short range interactions or long range interactions. In this paper, we will focus only on developing the SBC for a gas of hard disks.

This paper is organised as follows. In Section \ref{sec:2}, we describe the 
basic framework for this method, and how higher order algorithms can be implemented in general. In Section \ref{sec:3}, we restrict ourselves to a gaseous system of hard disks, and elaborate how this can be simulated using a first order algorithm. In the same section, we report various tests to ensure that the SBC is performing as we expect, before giving results of our calculations of the excess pressure, specific heat capacity, and chemical potential. We then conclude in Section \ref{sec:4}.

\begin{figure}[!ht]
\begin{center}
\includegraphics[scale=0.5]{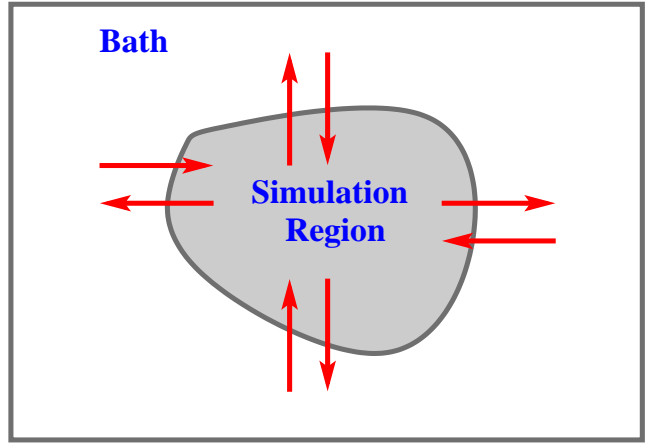}
\end{center}
\caption{For a finite simulation region inside an infinite heat bath, there is constant exchange of 
particles as well as momenta at the boundary.}
\label{figsys}
\end{figure}
\section{\label{sec:2}Overview and algorithms}
\subsection{Basic Notions}

\begin{figure}[!ht]
\begin{center}
\includegraphics[scale=0.4]{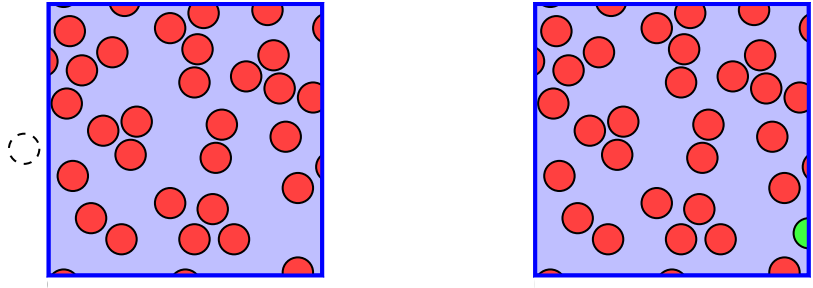}
\end{center}
\caption{In a gas of hard disks, particles are deleted when they are ejected (left, deleted particle shown as dashed circle) and new particles are injected at a boundary (right, injected particle shown in green).}
\label{gassbc}
\end{figure}
\begin{figure}[!ht]
\begin{center}
\includegraphics[scale=0.4]{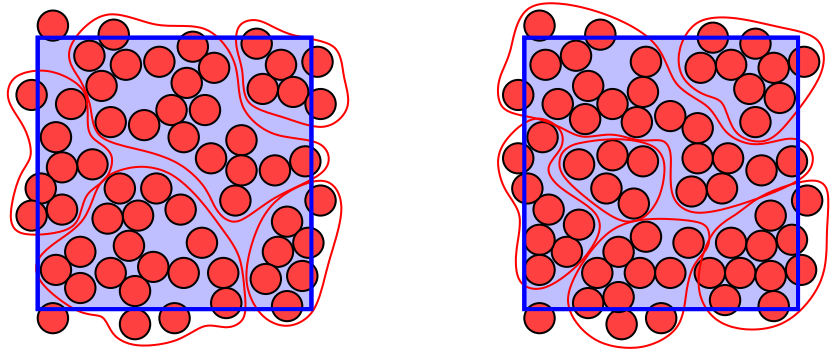}
\end{center}
\caption{In a liquid of hard disks, particles form dense clusters with varying particle numbers and shapes. The effective dynamics of the liquid consists of diffusion of the clusters, as well as particle and energy exchange between clusters.}
\label{liquidsbc}
\end{figure}
\begin{figure}[!ht]
\begin{center}
\includegraphics[scale=0.4]{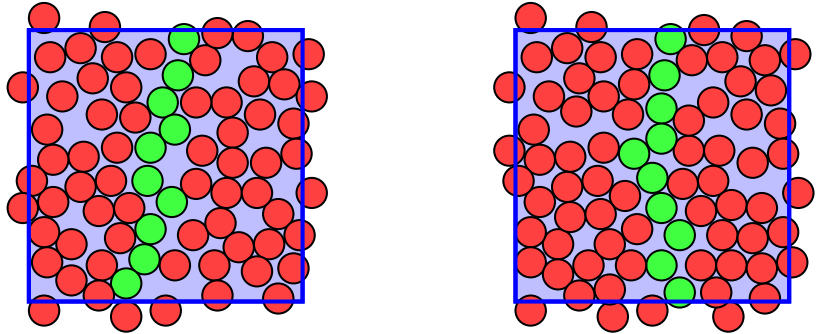}
\end{center}
\caption{The shape fluctuations of a monolayer of particles at the center of a hard disk solid. The fluctuation statistics of this layer is resampled to simulate the stochastic fluctuations at the boundaries.}
\label{solidsbc}
\end{figure}
Let us consider a finite observation volume 
embedded inside an infinite heat bath. At thermal equilibrium, 
the observed system exchanges particles with the environment.
This exchange of particles occurs only 
at the boundaries (see Figure \ref{figsys}). In addition, the system also exchanges energy with the environment, through the exchange of particles, as well as through long-range interactions with its fluctuating environment. The characters of these exchanges are qualitatively different in the gas, solid and liquid phases. In the gas and liquid phases, we have the phenomenon of diffusion. In these phases, we find exchange of particles at the boundaries, as well as the exchange of energy due to the exchange of particles. In the gaseous phase (see Figure \ref{gassbc}), ejection or injection involves only a single particle nearly all the time. In the liquid phase (see Figure \ref{liquidsbc}), particles form strongly-correlated clusters. Each cluster diffuses through the simulation system, and also exchanges particles with other clusters. Therefore, it is not accurate to have single-particle injections or ejections in such simulations. Instead, we should eject or inject clusters. In addition, the time evolutions of the shapes of the clusters must also be simulated as part of the SBC for liquids with short-range interactions. In the solid phase (see Figure \ref{solidsbc}), we have strong local spatial ordering of the particles. This greatly reduces the exchanges of particles. However, the system do exchange energies with the infinite system it embedded in through the collisions of particles at the boundaries. Thus simulating the fluctuations of the boundary particles without injecting or ejecting events should be sufficient for the SBC simulation of hard disk solids. For systems with long-range interactions, the dynamics of particles in the observation volume are affected by particles both inside and outside of the volume. To simulate such systems using SBC, the stochastic forces contributed by environmental particles must be simulated, again using a resampling method. In summary, we see that SBC, with appropriate resamplings, can be applied to systems with different densities and different interactions. 

\begin{figure}[!ht]
\begin{center}
\includegraphics[scale=0.4]{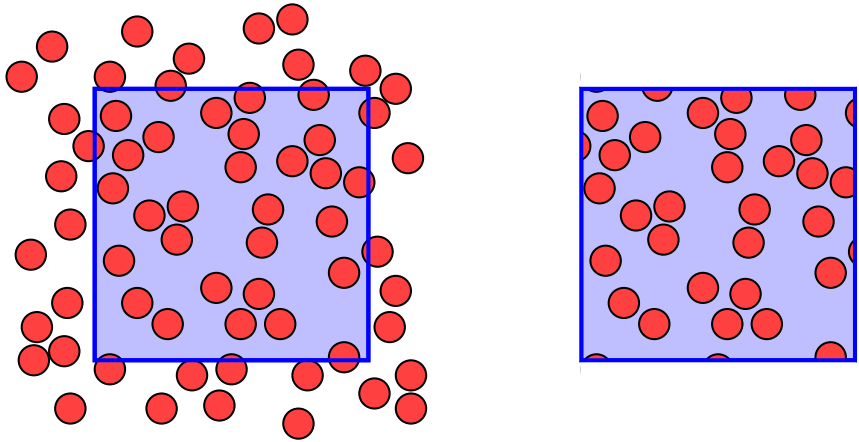}
\end{center}
\caption{A finite observation window within an infinite system (left) and a finite simulation system of the same size (right).}
\label{simsys}
\end{figure}
\subsection{Hierarchy of SBCs}
To better explain the resampling ideas behind the SBC, we will focus only on the simulation of a hard disk gas in this paper. For a finite observation volume inside an infinite system of hard disks (see Figure \ref{simsys}),
exchange of energy only occurs through the exchange of particles.  Therefore, if we could
somehow simulate the infinite system, we can record the history $\{\mathbf{r}_1,
\mathbf{r}_2, \dots, \mathbf{r}_i, \dots \}$ and $\{\mathbf{v}_1, \mathbf{v}_2,
\dots, \mathbf{v}_i, \dots \}$ of particle injected into the observation volume
at times $\{t_1, t_2, \dots, t_i, \dots\}$.  If we start with the same initial
conditions, and simulate the observation volume only, we would obtain the same
trajectories within the observation volume if we inject particles at times
$\{t_1, t_2, \dots, t_i, \dots\}$ with velocities $\{\mathbf{v}_1, \mathbf{v}_2,
\dots, \mathbf{v}_i, \dots\}$ at positions $\{\mathbf{r}_1, \mathbf{r}_2, \dots,
\mathbf{r}_i, \dots\}$ along the boundaries of the observation volume.  All
physical quantities that can be measured within the observation volume alone
would also have the same values in both simulations.

Of course, if we could obtain this history of particle injections from the infinite
system simulation, there would be no point running another simulation of the
finite observation volume.  The main idea behind our SBC method 
is to generate an artificial history of particle injections that
would mimick the presence of the infinite environment, without having to
simulate such an infinite environment.  To do so, we observe that for a
different initial condition, the hypothetical infinite system simulation would
produce a different history $\{\mathbf{r}'_1, \mathbf{r}'_2, \dots,
\mathbf{r}'_i, \dots\}$ and $\{\mathbf{v}'_1, \mathbf{v}'_2, \dots,
\mathbf{v}'_i, \dots\}$ at times $\{t'_1, t'_2, \dots, t'_i, \dots\}$.
Ultimately, these two particle injection histories (along with infinitely many
more) belong to the same statistical ensemble associated with the infinite
system being in thermal equilibrium at temperature $T$.  We therefore expect
them to be independent samples from the same fixed statistical distributions.

In principle, these particle injection history distributions that we must sample
from are very high dimensional, because they must incorporate all possible
spatial and temporal correlations between the entire history of particle
injections.  For simulation purposes, it is not feasible to work with these:
it is impossible to estimate the distributions from data, and also nearly
impossible to sample from them.  We must therefore always work with approximate
versions of these particle injection history distributions.  There is a
hierarchy of spatial and temporal approximations that we can make.  If we ignore
spatial correlations, and treat successive injections as statistically
independent, we can work with the much simpler distributions $f_0(\mathbf{r}_i)$
and $g_0(\mathbf{v}_i)$ for the injection positions and injection velocities
respectively.  These distributions can be derived theoretically, or estimated
from data. An algorithm to generate artificial particle injection histories
based on $f_0(\mathbf{r}_i)$ and $g_0(\mathbf{v}_i)$ would then be called
\emph{zeroth order in space}.  If we decide that spatial correlations are
important, and should not be completely ignored, we can work with the
conditional distributions $f_1(\mathbf{r}_i | \mathbf{r}_{i-1})$ and
$g_1(\mathbf{v}_i | \mathbf{v}_{i-1})$, where the probability of injecting a
particle at $\mathbf{r}_i$ with velocity $\mathbf{v}_i$ depends on where,
$\mathbf{r}_{i-1}$, and with what velocity, $\mathbf{v}_{i-1}$, the previous
particle had been injected.  An algorithm to generate artificial particle
injection histories based on $f_1(\mathbf{r}_i | \mathbf{r}_{i-1})$ and 
$g_1(\mathbf{v}_i | \mathbf{v}_{i-1})$ would be called \emph{first order in
space}.  In general, an algorithm that is $n$th order in space would be based on
the conditional distributions $f_n(\mathbf{r}_i | \mathbf{r}_{i-1}, \dots,
\mathbf{r}_{i-n})$ and $g_n(\mathbf{v}_i | \mathbf{v}_{i-1}, \dots,
\mathbf{v}_{i-n})$.  As we can imagine, higher-order conditional distributions
are hard to derive theoretically, and also hard to estimate from data.

Similarly, we can also design algorithms of various orders in time, depending on
what approximations we make, and what temporal correlations we ignore.  If we
inject particles into the observation volume at fixed time intervals of
$\bar{\tau} = \overline{t_i - t_{i-1}}$, the algorithm can be called
\emph{zeroth order in time}.  Alternatively, if we gather statistics for the
injection delays $\tau_i = t_i - t_{i-1}$, and then sample random delays from
this distribution, the algorithm can be called \emph{first order in time}.  In
general, in an algorithm that is \emph{$n$th order in time}, we will have to
gather statistics for $n$ successive injection (and perhaps also ejection)
events, and then have the random delays sampled from these conditional
distributions.  In this paper, we focus on developing a SBC for simulating a
hard disk gas.  Since the density of particles is low, we expect very little
spatial correlations between boundary events, and so we will stick to an
algorithm that is zeroth order in space.  However, the zeroth order in time
algorithm is too artificial, so we will develop instead an algorithm that is
first order in time.

\section{\label{sec:3}Application to a gaseous system of hard disks}
To test our first-order algorithm, we simulate a two-dimensional system of hard disks enclosed in an unit square. This simple system is chosen because there are no inter-particle potentials to deal with. Particle trajectories are therefore straight lines between collisions. For numerical simplicity, we set the Boltzmann constant $k_B$ and mass $m$ of the particles both to unity. We also fix the radius of the hard disk to be $\sigma = 0.005$. This leaves us with only two free parameters, the thermodynamic temperature $T_0$ and the density of particles $\eta = N_0\pi\sigma^2/V$ ($N_0$ is the number of particles and $V = 1$ is the volume of the system), that will determine the properties of the system.

\subsection{Simulation Procedure}
We simulated this system using the event-driven algorithm described by Alder and Wainwright~\cite{AW-59}, with $N_0$ particles. To speed up the computation we use the cell list scheme described in~\cite{AT-91,FS-02,QB-73}. Because ejection statistics must be collected before we can turn on SBC, we start the simulation off using PBC. Once enough data has been gathered, we replace the PBC by the SBC in our simulations. 

\subsubsection{Initialization}
We started the PBC simulation by assigning the $N_0$ particles random non-overlapping positions. We also gave each particle a random velocity sampled from the Maxwell-Boltzmann distribution at temperature $T_0 = 25$. Because the average velocity of the system is typically non-zero after the random assignment, we subtract the average velocity from all velocities $\{\mathbf{v}_i\}$, to obtain a set of velocities
\begin{equation}
\mathbf{v}_i' = \mathbf{v}_i-\dfrac{1}{N_0}\sum_i^{N_0}\mathbf{v}_i\,,
\end{equation}
whose average is zero. We then rescale components of the new set of velocites, $\{\mathbf{v}'_i\}$, independently, 
\begin{equation}
{\bf v_{i,x}^{new}} = {\bf v_{i,x}^{\prime}} \sqrt{\frac{N_0 \,T_0}{T_x}}, \quad 
{\bf v_{i,y}^{new}} = {\bf v_{i,y}^{\prime}} \sqrt{\frac{N_0 \,T_0}{T_y}}\,,
\end{equation}
Here, 
\begin{equation}
T_x = \sum_i^{N_0} {v_{i,x}^{\prime}}^2\quad {\rm and}\quad T_y = \sum_i^{N_0} {v_{i,y}^{\prime}}^2\,
\end{equation}
with $v_{i,x}'$ and $v_{i,y}'$ being the $x$ and $y$ components of the momenta $\mathbf{v}_i'$.
\subsubsection{Calibration}
\begin{figure}[!ht]
\begin{center}
\includegraphics[scale=0.3]{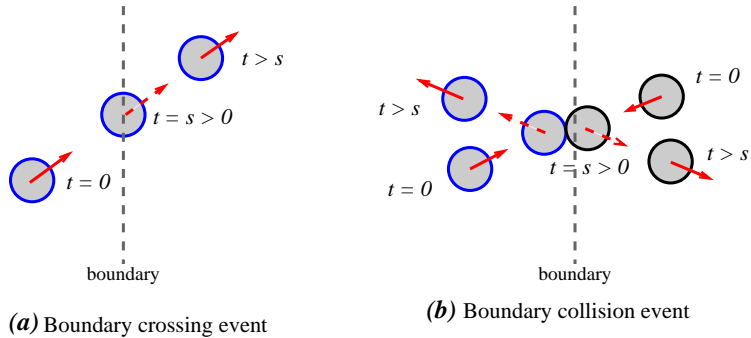}
\caption{The two classes of boundary events, (a) ejection, and (b) collision, 
considered for SBC statistics. During ejection a particle from inside the system leaves 
the system, whereas during a collision two particles on either side of the boundary 
scatter off each other.} 
\label{bevent}
\end{center}
\end{figure}

\begin{figure}[!ht]
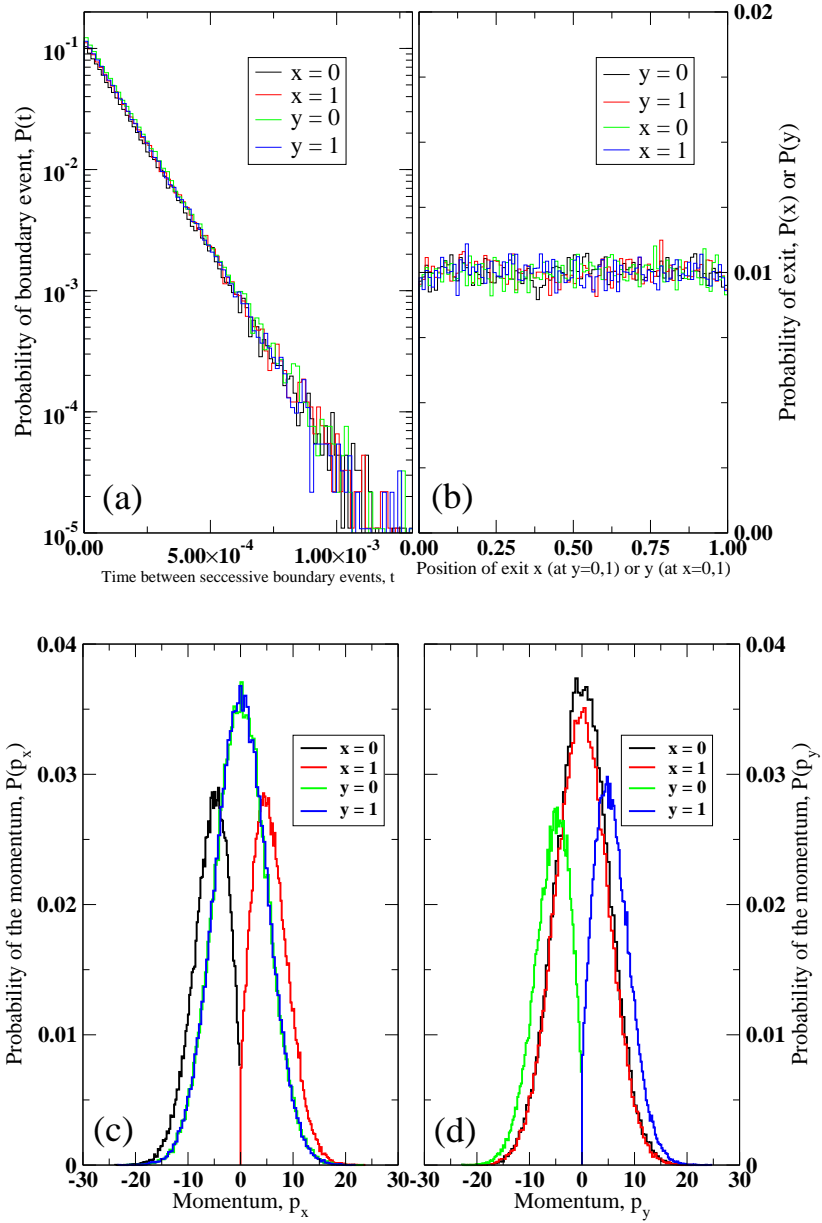

\begin{center}
\includegraphics[scale=0.40]{trdist.eps}
\label{xx}
\vspace{6mm}
\includegraphics[scale=0.40]{momdist.eps}
\end{center}
\caption{The distributions of exit events for each exit direction.
The delay time distribution of successive exit events is shown in (a), while (b) shows 
the distribution of positions at while they exit the system for each exit direction namely 
$x=0, x=1, y=0$ and $y=1$. In (c) and (d) we have plotted the distribution of the momentum components of the 
exiting particles. The figure shows that the exit statistics are independent of direction of exit and 
thus we treat injections in each direction as independent stochastic processes.}
\label{fig:dist}
\end{figure}

We then ran the simulation until there are at least 10000 exit events along each boundary. This is to ensure that the system has equilibrated from an initial condition that is not part of the equilibrium ensemble~\cite{H-92}. We then continued the PBC simulation until the number of ejections along each boundary is at least 75000. 
We fixed the number of exit events at 75000 to ensure that we have adequate statistics to estimate the time delay distribution. During this stage of the simulation, the distribution of time delays between successive boundary events (which includes both boundary crossings and collisions, as shown in Figure \ref{bevent}), the position and the momentum of the event particles are measured for each direction of boundary. 
These distributions are shown in Figure \ref{fig:dist}. From Figures \ref{fig:dist}(a) and \ref{fig:dist}(b), we observe that the boundary events occur uniformly along the boundary of the system while the time between successive events along each direction of boundary is independent of direction, as expected from the translational and rotational symmetries of the system. Since we have decided to keep our SBC algorithm zeroth order in space, this means that there is no need to keep the empirical distribution of injection positions. However, for our first order in time algorithm, we will need an empirical distribution of injection time delays, which should be the same as the distribution of ejection time delays because Newtonian mechanics is time reversal invariant. To obtain better statistics for this empirical distribution, we pool time delays collected from all four boundaries. 

Figures \ref{fig:dist}(c) and \ref{fig:dist}(d) show the momentum distribution of particles exiting the four boundaries. As expected, the component of the momentum parallel to the boundary follows a Gaussian distribution with mean zero. We expect the component of momentum normal to the boundary to follow a truncated Gaussian distribution of the form 
\begin{equation}
g_0(v_{\perp}) = \begin{cases} 
\sqrt{\frac{2}{\pi T}}\exp \left(-\frac{v_{\perp}^2}{2T} \right), & v_{\perp} \geq 0; \\
0, & v_{\perp} < 0. \end{cases}
\end{equation}
From the list of boundary events recorded, we can also extract the distribution of times between successive collisions at the boundary. Knowing this distribution of time between successive boundary collisions help us determine when a particle approaching the boundary will next collide, given the most recent collision at the given boundary. 
\begin{figure}[!ht]
\begin{center}
\includegraphics[scale=0.38]{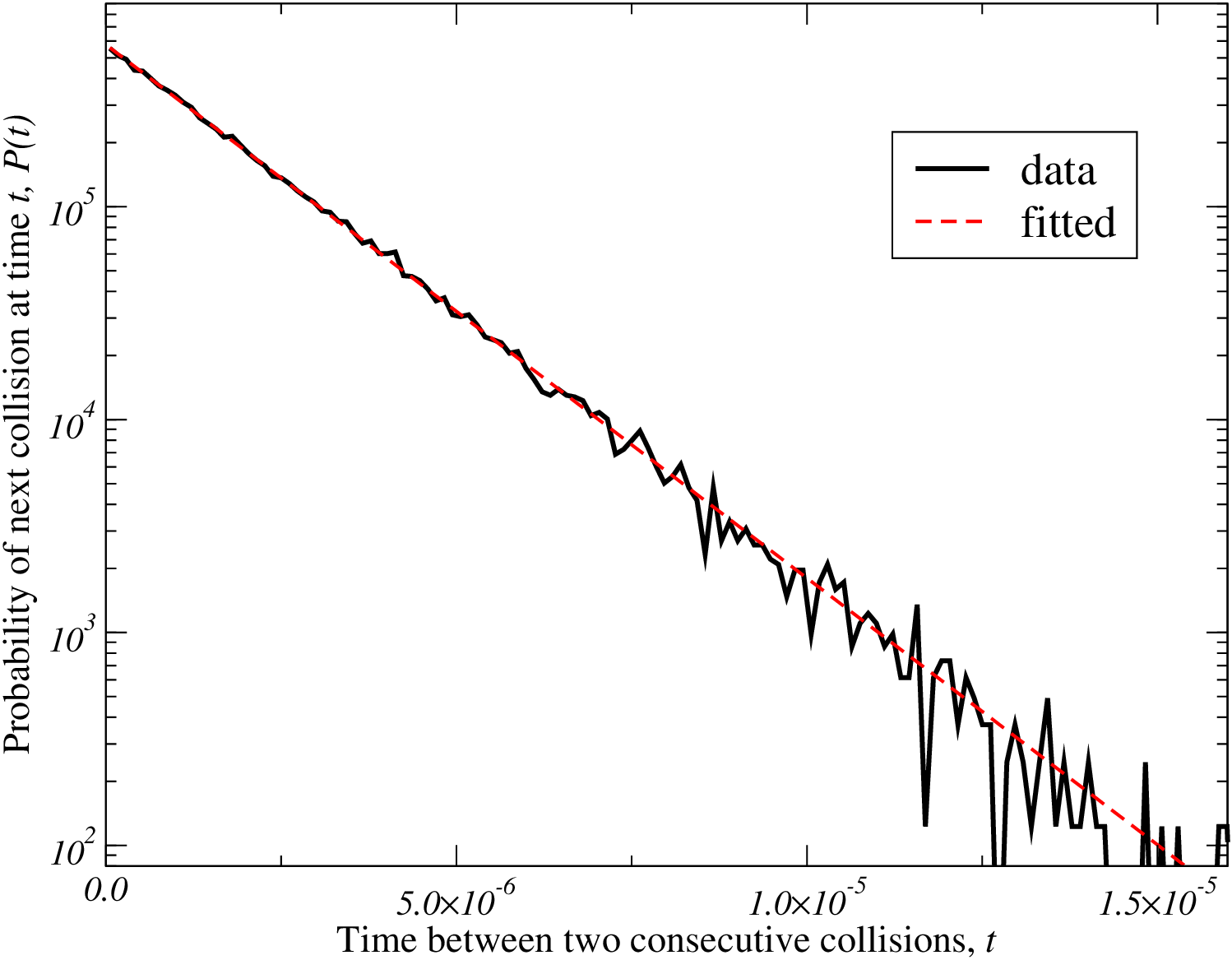}
\caption{Probability distribution of collision times for $N_0 = 2200$ at temperaure $T_0=25$. 
The equation used for the fitted curve (red dashed) is $f(t)=(1/\tau)\,\exp(-t/\tau)$, where 
$\tau = 1.7334\times10^{-6} $ is the mean time between any two collisions.}
\label{ctime}
\vspace*{4mm}
\includegraphics[scale=0.38]{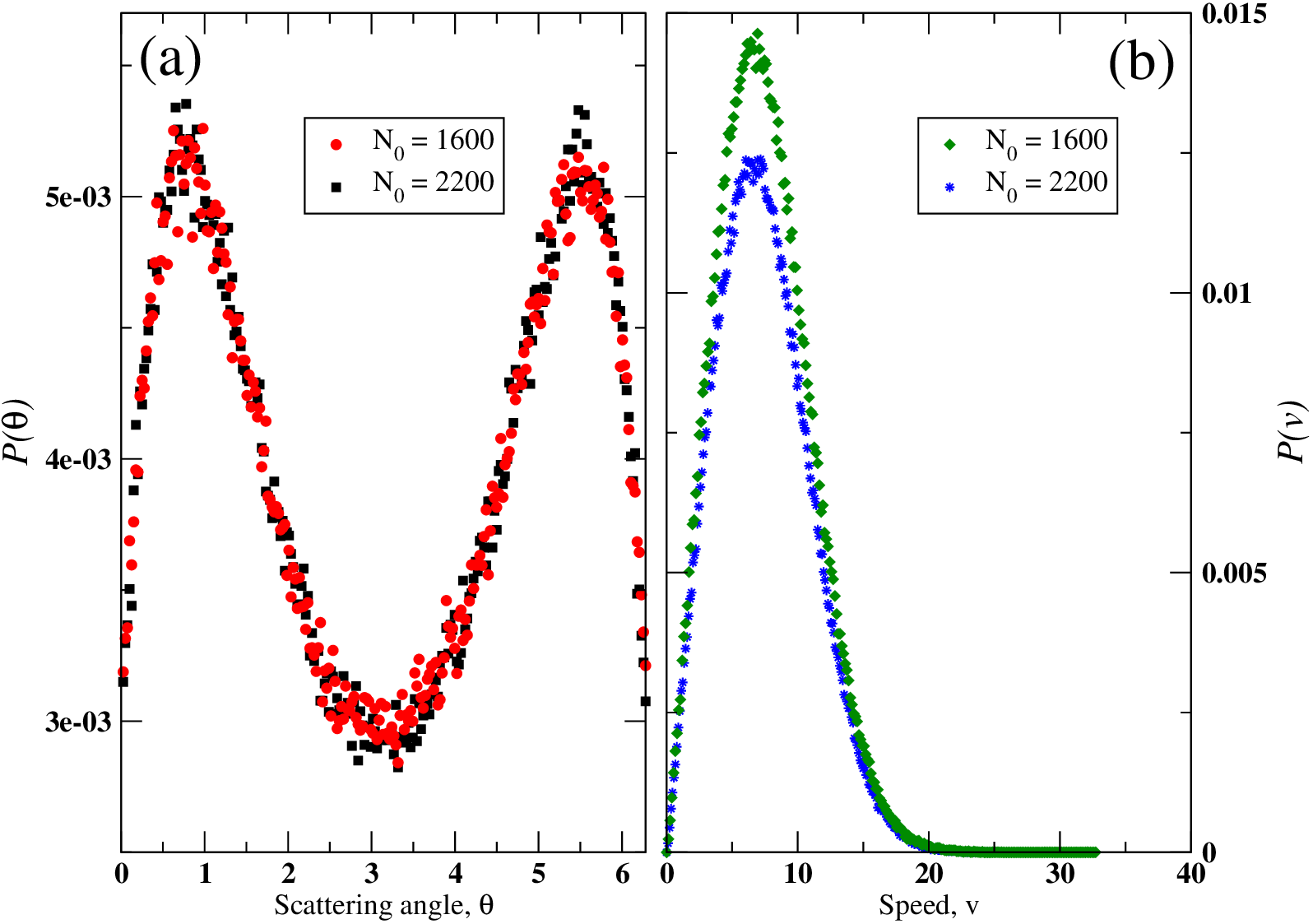}
\end{center}
\caption{The probability (a) $P(\theta)$ for the scattering angle measured relative to the incident angle of the particle and (b) $P(v)$ for the speed of the scattered particle for $N_0 = 1600$ 
and $N_0 = 2200$. The speed distribution
is consistent with a Gaussian initial velocity distribution. The 
temperature of the system is the same as in Figure \ref{ctime}.}
\label{stv}
\end{figure}
In order to simulate collisions with environmental particles at the boundaries, we also collect scattering angle and momentum transfer statistics. These distributions are shown in Figures \ref{ctime} and \ref{stv} respectively. 
  
\subsubsection{SBC Simulation}
Once the necessary distributions are calibrated, they are stored as histograms, before we turn off PBC and turn on SBC. In this SBC stage of the simulation, a system particle reaching the boundary will not always be ejected. Instead, we calculate the probability for it to be scattered at the boundary, based on the time of the last collision on that boundary. To make this more concrete, let $t_2$ be the time at which the particle reaches the boundary, and $t_1 < t_2$ be the time of the last collision at this boundary, which can be an attempted ejection or an attempted injection. The probability that this particle exits the system without undergoing any collision is 
\begin{equation}
P(t_2,t_1) = \exp\left[-\dfrac{(t_2-t_1)}{\tau}\right]\,,
\label{eqn:collprob}
\end{equation}
where $\tau$ is the average time between consecutive collisions. If we draw a $U(0,1)$ random number smaller than $P(t_2,t_1)$, the particle exits the system (and is thereafter deleted from the simulation). Otherwise, the particle undergoes a collision at the boundary, and is assigned a new velocity re-sampled from the scattering statistics (see Figure \ref{stv}). Depending on the scattered velocity, this particle either makes its way out, or is deflected back into the system. After each event in this SBC stage of the simulation, we update the histograms of the relevant distributions.

In the mean time, particles are injected into the system. This is done by first drawing a set of future injection times for each boundary, based on the statistic in Figure \ref{fig:dist}(a). Since this statistic is collected for both boundary crossing and boundary collision events, the injection events sampled from this statistic can also succeed or fail. When the next injection time $t_2$ is reached, we then determine whether it would be successful by comparing a $U(0,1)$ random number against $P(t_2,t_1)$ in Equation (\ref{eqn:collprob}), as we did for attempted ejections. If the injection is successful, the injected particle will be assigned a random velocity sampled from the distributions in Figures \ref{fig:dist}(c) and \ref{fig:dist}(d). Since Figure \ref{fig:dist}(b) shows that the boundary crossing events should be uniformly distributed along the boundaries, the successful injection is carried out at a random empty space along the boundary.

\subsubsection{Reversibility}
At thermodynamic equilibrium, we must have detailed balance. This means that within the equilibrium ensemble, the transition rate from one state to another must be equal to the transition rate from latter to the former~\cite{HB-85, KP-98}. Instead of an exhaustive survey of the high-dimensional phase space, we demonstrate that our SBC simulations indeed satisfy detailed balance by running time reversed versions of these simulations. To simulate the time reversed system, all velocities inside the system were reversed after a fixed duration $t$ of SBC simulation. Thereafter, the time reversed SBC system was simulated also for duration $t$. In all the testing results presented in Subsection \ref{sec:3a}, we always compare the SBC distributions against their time-reversed analogs. 

\subsection{\label{sec:3a}Testing}
\subsubsection{Equilibrium}
\begin{figure}[!ht]
\begin{center}
\includegraphics[scale=0.45]{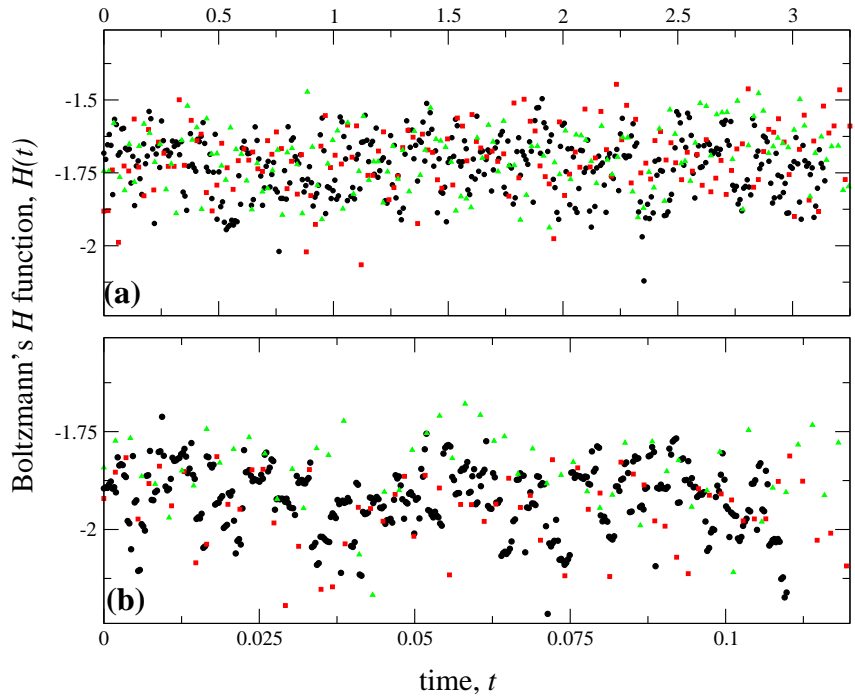}
\end{center}
\caption{Comparison of Boltzmann's $H$-fucntion as a function of time calculated both from PBC (black) 
simulation against the SBC simulation (colored), for different initial densities (a) $N_0 = 500$ and 
(b) $N_0=3000$. The graphs show that the equilibrium attained during the PBC simulation is preserved by the SBC simulation.}
\label{bmanh} 
\end{figure}

Before we compute any thermodynamic variables we must first ensure that the simulation system has equilibrated, and its total energy and entropy remain more or less constant. We check this by first computing the system's Boltzmann's $H$-function~\cite{H-92} as a function of time, in place of the entropy of the system. In Figure \ref{bmanh}, we compare the results of SBC simulations against a PBC simulation. We see that SBC preserves the thermodynamic equilibrium attained during the PBC stage of our simulation. 

\subsubsection{Detailed balance}
\begin{figure}[!ht]
\begin{center}
\includegraphics[scale=0.4]{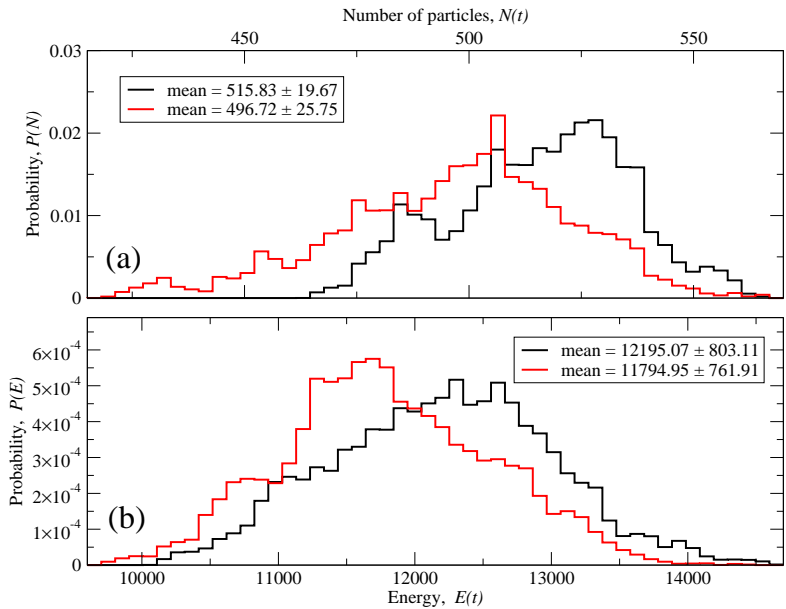}
\caption{The comparison of statistics of (a) particle number and (b) total energy of the system for the SBC simulated forward in time (black) and simulated backwards in time (red) for $N_0=500$. The simulation time for this graph is 
$t=4$. We observe that the time reversed distributions are within one standard deviation of their forward-time counterparts.} 
\label{timerev}
\end{center}
\end{figure}

To check the more stringent condition of detailed balance for thermodynamic equilibrium, we compare the statistics of particle number and total energy of the forward time SBC and the time reversed SBC simulations for $N_0 = 500$ (see Figure \ref{timerev}).
While simulating the SBC forward in time, the statistics used are mostly from the PBC stage of simulation. On the other hand, when simulating the SBC backward in time, the statistics used receive more contribution from the SBC simulation running forward in time. Because of this systematic difference, small discrepancies are expected in the distributions. However, looking at the figures we can safely say that the SBC algorithm satisfies the detailed balance condition. 

\subsubsection{Particle number and average energy}
\begin{figure}[!htbp]
\begin{center}
\includegraphics[scale=0.35]{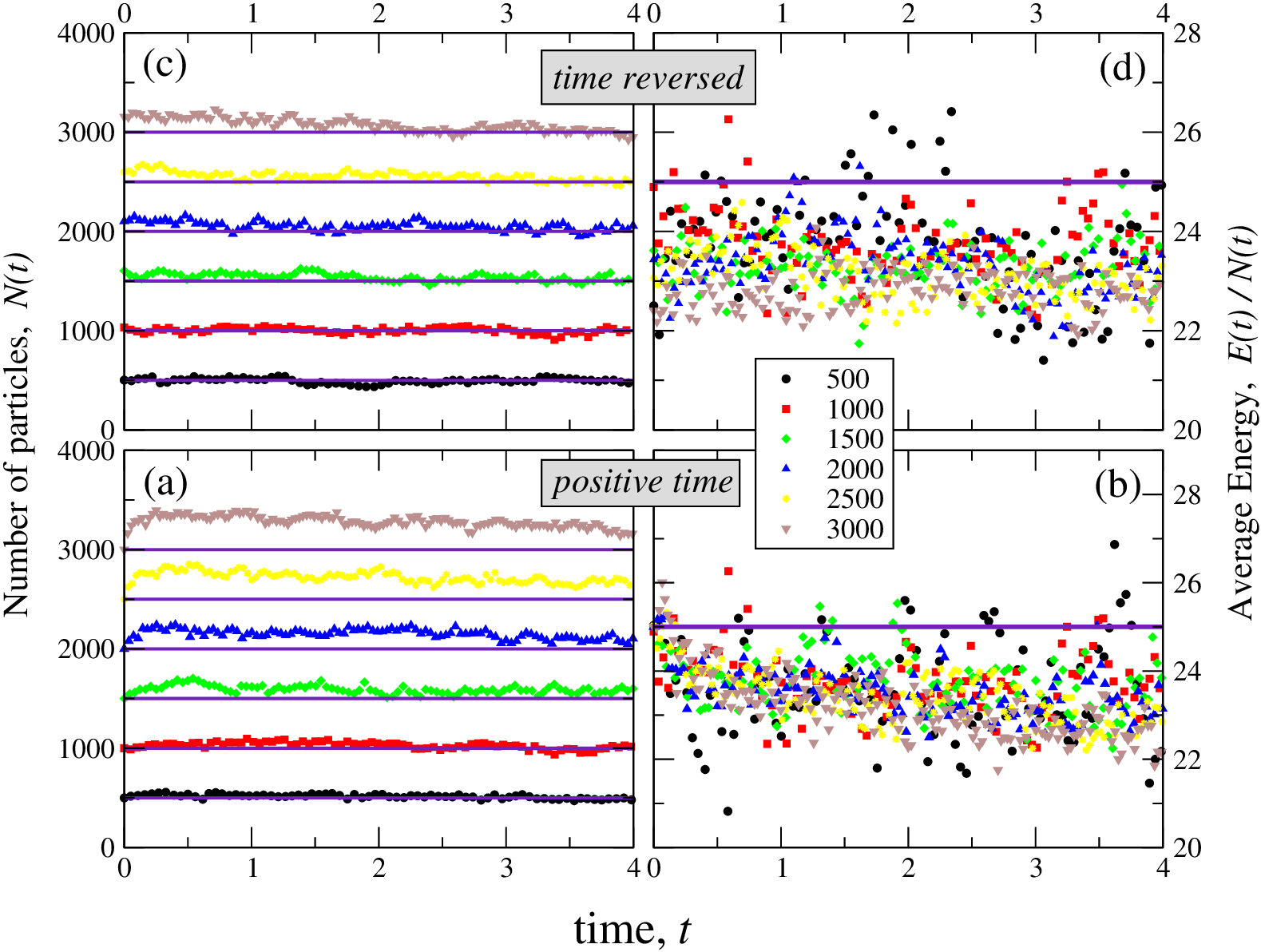}
\caption{The variation of (a) the particle number of the system and (b) the average 
energy during the SBC simulation as a function of time for different densities ($N_0=500, 1000, 1500, 
2000, 2500$ and $3000$). The graphs (c) and (d) show the same variations but for time 
reversed system (see text for details). For low densities the number of particles inside the system 
fluctuates about the initial particle number $N_0$ (straight line) whereas for all the simulations the 
total energy inside the system remains near to the initial energy (straight line, calculated through
 the equipartition theorem at $T_0=25$).}
\label{partene} 
\vspace*{3mm}
\includegraphics[scale=0.35]{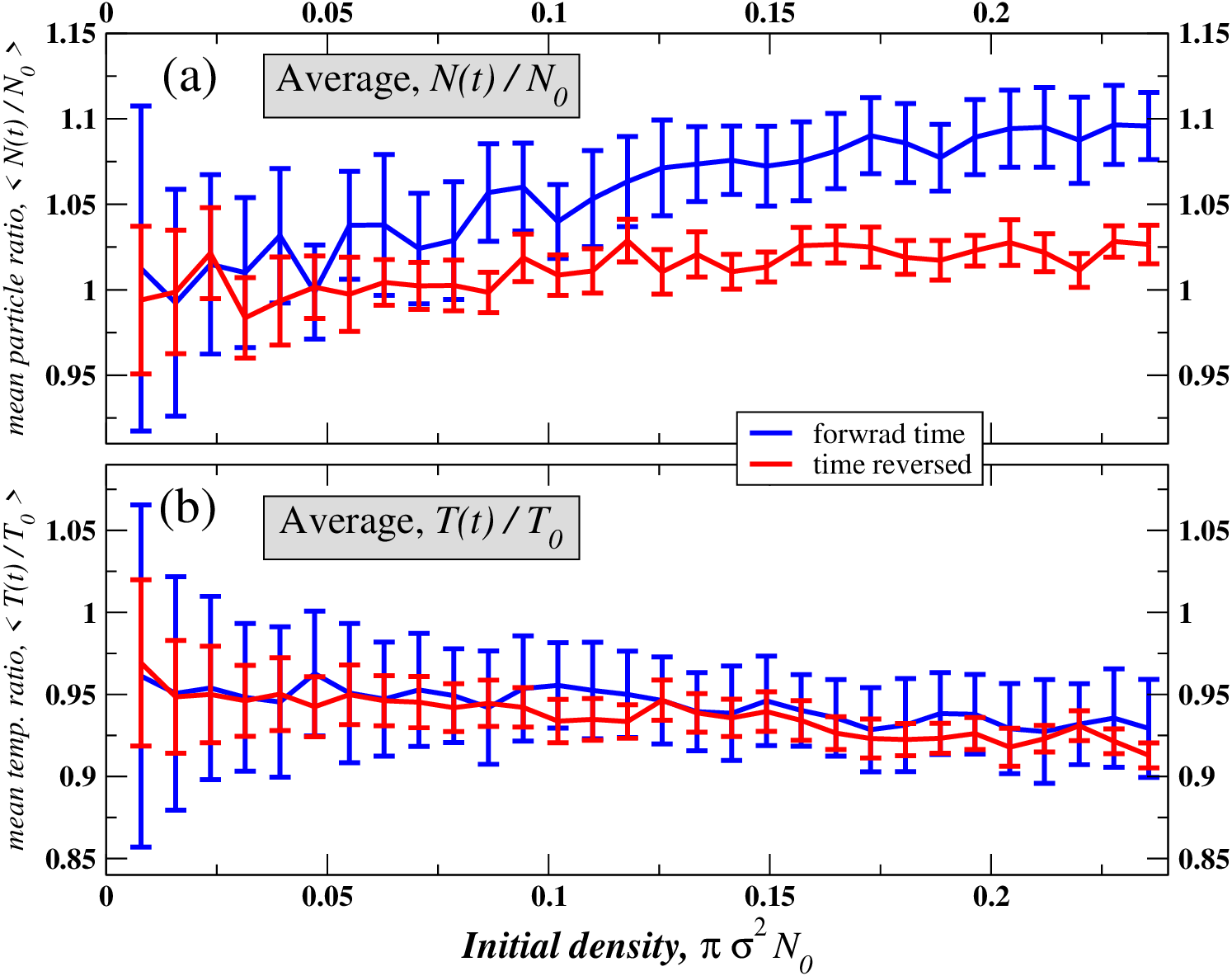}
\caption{The variation of (a) the mean number of particles inside the system and (b) the mean particle  
temperature during the SBC simulation as a function of initial density of the simulation. The temperature 
has been calculated from the equipartition theorem. The blue curve is for positive time simulation where 
as the red curve is for the time reversed system.}
\label{parteneden} 
\end{center}
\end{figure}

After checking that the system is at equilibrium and detailed balance is maintained within the system, we compute the particle number and the total energy of the system after each injection (or ejection) of a particle into the system (or out of the system). We compare these data of number of particles inside the system and the energy per particle against the initial particle number and total energy respectively. This is shown in Figures \ref{partene} and \ref{parteneden} for various initial densities. 
Within any given time interval, the number of injections may not be same as the number of ejections. Thus we expect, the particle number $N$ as well as the total energy $E$ to fluctuate with time. For low densities, we can see that the particle number inside the system fluctuates about the initial number of particles. But for higher densities, we observe small discrepancies shortly after SBC was turned on. These are expected for the following reason. When PBC was imposed the simulation builds up excess correlations that get incorporated into the distribution functions. Once we turn on SBC these correlations create an imbalance between injection and ejection rates. However, the simulations eventually relaxes to an SBC equilibrium with the correct average $N$ and $E$. In some cases, this relaxation is slow. We expect that higher order SBC algorithms described in Section 2 will speed up this relaxation from the PBC equilibrium. 

From equipartition theorem the kinetic energy per particle should be equal to the temperature of the system (with Boltzmann's constant $k_B=1$). We find from Figure \ref{partene} that the energy remains close to this value. We believe the most important reason for the small discrepancy is the first order algorithm missing out correlations due to the finite particle size. Because of the finite particle size, the interaction between the bath and the system happens in a finite region about the boundary (extending up to the radius of the particle in both directions from the boundary). To faithfully reproduce the correlations within this interaction zone, we have to go to higher order algorithms. As the density increases further, we will find more multiple collisions within the interaction zone. 

\subsubsection{Pair distribution function}
\begin{figure}
\begin{center}
\includegraphics[scale=0.4]{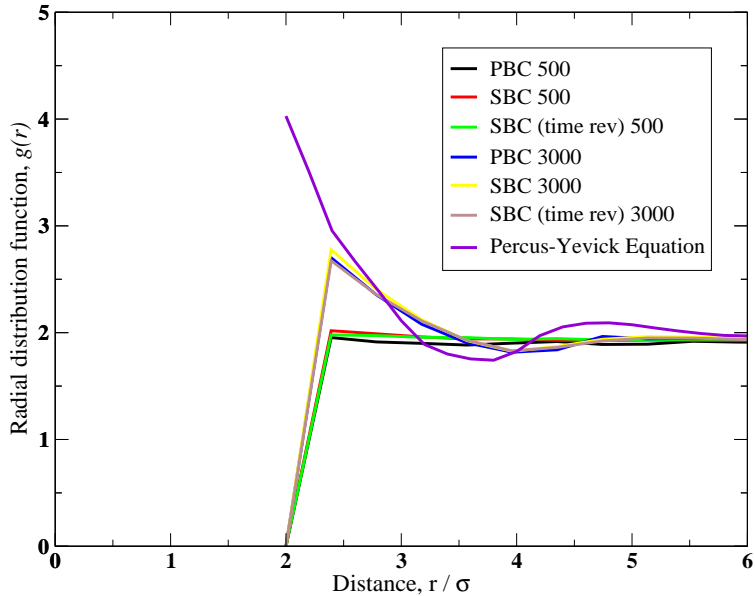}
\end{center}  
\caption{Comparison of radial distribution functions for the PBC and SBC MD, together with the values calculated from Percus-Yevick equation \cite{DFT-69}. The graphs are for two initial
densities $N_0=500$ and $N_0=3000$ of the simulation. The Percus-Yevick curve is calculated at $\rho/\rho_0 = 0.4$, which is equivalent to $N_0 \approx 4619$. Note the similarity between the curve for $N_0 = 3000$ and that of the Percus-Yevick equation. The time reversed SBC MD are also plotted for the simulations.}
\label{raddist}
\end{figure}

The pair distribution $g(r)$ function gives the probability of finding a pair of atoms a distance $r$ apart. It not only provides insight into the average structure of the system, but is also useful for calculating ensemble average e.g., energy, pressure as well as the chemical potetial~\cite{AT-91}. To calculate $g(r)$, we first measure the separations between all pairs of atoms in a given configuration. These are then binned into histograms, $n(r)$, which provides the relative number of atoms between a distance $r$ and $r+\delta r$, $\delta r$being the bin size for the histograms. The radial distribution function $\rho g(r)$is then calculated by using 
\begin{equation}
\rho g(r) = \dfrac{n(r)}{2\pi r\delta r}\,,
\end{equation}
$\rho$ being the density of the system, averaging over 10,000 equilibrium configurations. 
In Figure \ref{raddist} we compare $g(r)$ calculated from PBC and SBC simulations, for $N_0 = 500$ and $N_0 = 3000$. As we can see, the PBC and SBC results are in good agreement. At $N_0 = 3000$, oscillatory features can be seen in $g(r)$, telling us that we are already near liquid density. Beyond this density, long range correlations cannot be ignored in simulations, and even the higher order versions of the gas algorithm described in this paper may not be accurate enough. Instead, we will need to develop SBC algorithms specifically for liquid densities, as described in Section \ref{sec:2}.

\clearpage

\subsubsection{Compressibility}
\begin{figure}[!htbp]
\begin{center}
\includegraphics[scale=0.38]{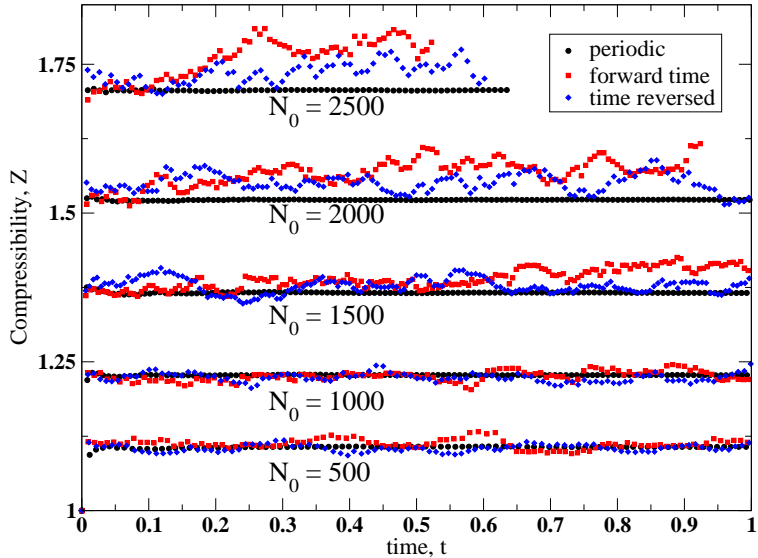}
\caption{The measured compressibility as a function of time for various initial densities (colored) are shown 
in comparison to the PBC results (black). It shows that for very small densities the value of compressibility 
is consistent with the PBC simulation. At higher densities, we find persistent discrepancies between SBC and PBC compressibilities.}
\label{press}
\end{center}
\end{figure}

For hard disk simulations, compressibility of the system can be computed from collisions within the system~\cite{H-92}
\begin{equation}
Z = \dfrac{PV}{Nk_BT} = 1+\dfrac{2m\sigma}{3E_k}\dfrac{1}{t}\sum^{N_c}_{c=1}|\Delta v_{ij}(t_c)|\,,
\end{equation}
where $N_c$ is the total number of collisions occurring up to time $t$ and $\Delta v_{ij}(t_c)$ is the change in velocities for the colliding particles $i$ and $j$. In Figure \ref{press}, we show the PBC and SBC compressibilities for different densities. At low densities, the PBC and SBC compressibilities agree. At high densities, the SBC system is more compressible than the PBC system. 

\subsubsection{Specific heat}
\begin{figure}[!ht]
\begin{center}
\includegraphics[scale=0.3]{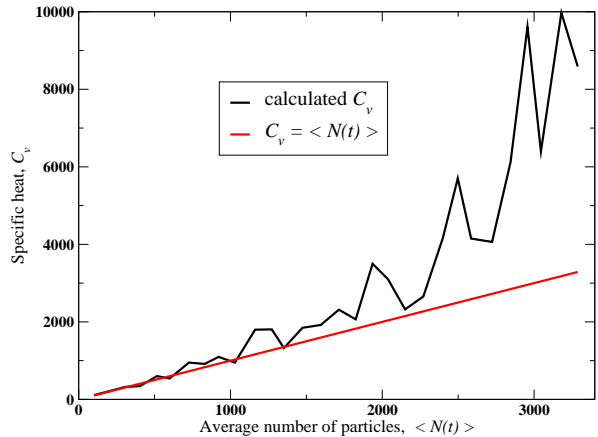} 
\end{center}
\caption{Variation of specific heat capacity $C_v$ with the average number of particles $\langle N(t)\rangle$
inside the system.}
\label{spc}
\end{figure}

Since the total energy and particle number of the system fluctuates with time in a SBC simulation, it is now possible to compute the specific heat
\begin{equation}
k_BT^2C_v = \left \langle (E-\langle E\rangle )^2\right \rangle -\dfrac{\left(\langle NE\rangle -\langle N\rangle \langle E\rangle \right )^2}{\left \langle (N-\langle N\rangle )^2\right \rangle}\,
\end{equation}
from the variance of fluctuations of these quantities~\cite{RKP-96}. In Figure \ref{spc} we show the specific heat as a function of the particle number. 
For an ideal 2D gas, the specific heat should be $C_v = Nk_B$. Our results are in good agreement with this theoretical specific heat for low densities, $N_0 < 2000$. At higher densities, our specific heat devaites strongly from the ideal gas behavior. Since the hard disk system at the highest density $N_0 = 3000$ in our simulations is already close to being a liquid, we expect the sharp rise in our specific heat around $N_0 = 3000$ to be a signature of the gas-to-liquid transition beyond this density.

\section{\label{sec:4}Conclusions}
To summarize, we described in this paper the basic framework for the molecular dynamics simulations of a finite system embedded in an infinite environment, by imposing stochastic boundary conditions that mimick the exchange of energy and particles between the system and its environment. In this method, particles that leave the system are deleted from the simulation, instead of being reinjected into the system through periodic boundaries. In addition, new particles are injected with random velocities at random positions along the boundaries at random times. In order to simulate the grand canonical ensemble, we chose the injection statistics to be time-reversed versions of the ejection statistics, which can be measured empirically by starting the simulation off in PBC. 

Applying this method to a hard disk gas, we showed by measuring the Boltzmann $H$-function, the particle number $N$, the average energy $E/N$, that the equilibrium attained the PBC stage of the simulation is preserved by our SBC. Like the PBC, we demonstrated that our SBC is also time-reversal invariant. We then proceeded to measure the pair distribution function $g(r)$, the compressibility $Z$, and the specific heat $C_v$ of the hard disk gas, and find that they agree with analytical results (where they are available). 

While we performed an equilibrium simulation in this paper, we expect our method to be useful for non-equilibrium situations as well. In particular we can choose any number of boundaries as control surfaces. At these control surfaces, we can fix different control parameters e.g., they can be at different temperature, or we can implement different injection velocity distributions or different rate of injection of particles. Such a method can be used for various application, such as (but not limited to) MD measurements f short time heat transport coefficients, jet intrusion, melting of solids by plasma as well as effusion through nanopores.

\subsection*{Acknowledgements}
This research work is supported by startup grant RG 19/07 and Academic Research Fund Tier 1 grant RG 22/09 from the Nanyang Technological
University.

\end{document}